\documentclass[aps,prl,twocolumn,showpacs,superscriptaddress]{revtex4}

\usepackage{graphicx}
\usepackage{amsmath}

\begin{document}

\title{Strong ferromagnetic exchange interaction in the parent state of the superconductivity in BaFe$_2$S$_3$}
\author{Meng Wang}
\email{wangm@berkeley.edu}
\affiliation{Department of Physics, University of California, Berkeley, California 94720, USA }
\author{S. J. Jin}
\affiliation{School of Physics, Sun Yat-Sen University, Guangzhou 510275, China }
\author{Ming Yi}
\affiliation{Department of Physics, University of California, Berkeley, California 94720, USA }
\author{Yu Song}
\affiliation{Department of Physics and Astronomy, Rice University, Houston, Texas 77005, USA }
\author{H. C. Jiang}
\affiliation{Stanford Institute for Materials and Energy Sciences, SLAC National Accelerator Laboratory and Stanford University, Menlo Park, California 94025, USA }
\author{W. L. Zhang}
\affiliation{Beijing National Laboratory for Condensed Matter Physics, Institute of Physics, Chinese Academy of Sciences, Beijing 100190, China }
\author{H. Q. Luo}
\affiliation{Beijing National Laboratory for Condensed Matter Physics, Institute of Physics, Chinese Academy of Sciences, Beijing 100190, China }
\author{A. D. Christianson}
\affiliation{Quantum Condensed Matter Division, Oak Ridge National Laboratory, Oak Ridge, Tennessee 37831, USA}
\author{E. Bourret-Courchesne}
\affiliation{Materials Science Division, Lawrence Berkeley National Laboratory, Berkeley, California 94720, USA }
\author{D. H. Lee}
\affiliation{Department of Physics, University of California, Berkeley, California 94720, USA }
\affiliation{Materials Science Division, Lawrence Berkeley National Laboratory, Berkeley, California 94720, USA }
\author{Dao-Xin Yao}
\affiliation{School of Physics, Sun Yat-Sen University, Guangzhou 510275, China }
\author{R. J. Birgeneau}
\affiliation{Department of Physics, University of California, Berkeley, California 94720, USA }
\affiliation{Materials Science Division, Lawrence Berkeley National Laboratory, Berkeley, California 94720, USA }
\affiliation{Department of Materials Science and Engineering, University of California, Berkeley, California 94720, USA }

\begin{abstract}

Inelastic neutron scattering measurements have been performed to investigate the spin waves of the quasi-one-dimensional antiferromagnetic ladder compound BaFe$_2$S$_3$, where a superconducting transition was observed under pressure [H. Takahashi {\it et al.}, Nat. Mater. 14, 1008-1012 (2015); T. Yamauchi {\it et al.}, Phys. Rev. Lett. 115, 246402 (2015)]. By fitting the spherically averaged experimental data collected on a powder sample to a Heisenberg Hamiltonian, we find that the one-dimensional antiferromagnetic ladder exhibits a strong nearest neighbor ferromagnetic exchange interaction ($SJ_R=-71\pm4$ meV) along the rung direction, an antiferromagnetic $SJ_L=49\pm3$ meV along the leg direction and a ferromagnetic $SJ_2=-15\pm2$ meV along the diagonal direction. Our data demonstrate that the antiferromagnetic spin excitations are a common characteristic for the iron-based superconductors, while specific relative values for the exchange interactions do not appear to be unique for the parent states of the superconducting materials.

\end{abstract}

\pacs{75.30.Ds,75.30.Et,78.70.Nx} 
\maketitle

The mechanism of high temperature ($HTC$) superconductivity has been one of the most intensely investigated topics since the discovery of the copper-oxide superconductors\cite{Si2016}. Analogous to the role of phonons in promoting superconductivity in conventional superconductors, spin fluctuations have been viewed as a possible glue that is essential for the formation of cooper pairs in the $HTC$ superconductors\cite{Bardeen1957,Scalapino2012}. It has been shown that the spin fluctuations in both copper and iron-based superconductors (FeSC) are intimately coupled with the superconductivity, specifically, the appearance of a spin resonance mode in the superconducting (SC) state, and the doping dependence of the spin fluctuations in the normal state\cite{Dai2015,Wangm2013}. The spin fluctuations in a SC compound derive from the spin waves of its magnetically ordered parent compound. Measurements of the spin waves in the parent compound are essential to determine the nature of the spin fluctuations and, in turn, to elucidate their role in the $HTC$ superconductors including the possibility that the spin fluctuations are the primary pairing mechanism.

\begin{figure}[t]
\includegraphics[scale=0.5]{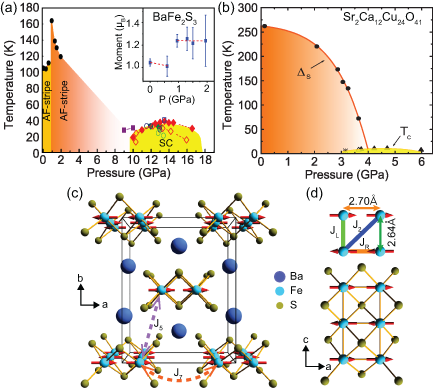}
\caption{ (a) Pressure dependence of the AF and superconducting transitions, and the moment sizes (inset panel) for BaFe$_2$S$_3$ adopted from Ref. [\onlinecite{Chi2016}]. (b) Pressure dependence of the spin gap ($\Delta_s$) and superconducting transitions for the laddered compound Sr$_{2}$Ca$_{12}$Cu$_{24}$O$_{41}$ adopted from Ref. [\onlinecite{Piskunov2001}]. (c) A sketch of the ladder structure of BaFe$_2$S$_3$. The cuboid indicates one unit cell. (d) One-dimensional edge-shared FeS tetrahedra in BaFe$_2$S$_3$. The red arrows represent the moment directions of irons. The $J_L, J_R, J_2, J_5$ and $J_7$ are the magnetic exchange interactions between the corresponding irons. }
\label{fig1}
\end{figure}

\begin{figure*}[t]
\includegraphics[scale=0.6]{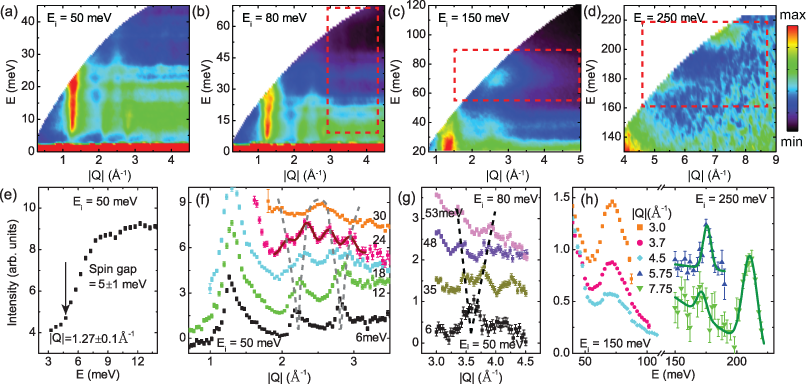}
\caption{ (a) INS spectra $S(Q,\omega)$ of BaFe$_2$S$_3$ at 5 K with $E_i=50$, (b) 80, (c) 150, and (d) 250 meV. The color represents intensities in arbitrary units. The red dashed rectangles highlight the areas for the cuts in (g) and (h). (e) Constant $Q$ cut with $E_i=50$ meV between $1.26<Q<1.28$ \AA $^{-1}$. (f) Constant energy cuts at $E=6$, 12, 18, 24, and 30 meV integrated within $E\pm1$ meV with $E_i=50$ meV. (g) Similar constant energy cuts at $E=6\pm1, 35\pm1$ meV with $E_i=50$ meV and $E=48\pm1.5, 53\pm1.5$ meV with $E_i=80 meV$ at 5 K.  The dashed lines are guides to the dispersion relations of spin excitations. The solid line on top of $E=24$ meV data points is a fit to Gaussian functions. The intensities for $E=24$, 48, and 53 meV have been doubled for comparison. (h) Constant $Q$ cuts at $Q=3.0$, 3.7, 4.5 \AA $^{-1}$ integrated within $Q\pm0.15$ \AA $^{-1}$ with $E_i=150$ meV and $Q=5.75$, 7.75 \AA $^{-1}$ integrated within $Q\pm0.25$ \AA $^{-1}$ with $E_i=250$ meV. The green solid lines are fits to Gaussian functions. The error bars are one standard deviation of the measured counts.}
\label{fig2}
\end{figure*}

Recently, a SC transition up to 24 K has been observed in the quasi-one-dimensional (1D) ladder compound BaFe$_2$S$_3$ under pressure in the range of 10 to 17 GPa\cite{Takahashi2015,Yamauchi2015}. The obtained pressure dependent phase diagram (Fig. \ref{fig1} (a)) resembles that of the 1D copper oxide laddered system Sr$_{14-x}$Ca$_{x}$Cu$_{24}$O$_{41}$ (Fig. \ref{fig1} (b))\cite{Uehara1996,Nagata1999,Piskunov2001,Vuletic2006} and the commonly observed doping dependent phase diagrams in the layered FeSC\cite{Stewart2011}. This suggests that BaFe$_2$S$_3$ at ambient pressure is the parent state of the superconductivity discovered under pressure, and that the superconductivity likely has a common origin, possibly magnetic-fluctuation-mediated\cite{Arita2015}. It has been suggested that the abrupt increase of the N$\acute{e}$el temperature ($T_N$) as a function of pressure shown in Fig. \ref{fig1} (a) is associated with a quantum phase transition due to the change of orbital occupancies under pressure\cite{Chi2016}. BaFe$_2$S$_3$ is isostructural with the 1D antiferromagnetic (AF) ladder compounds $A$Fe$_2$Se$_3$ ($A=$ K, Rb, and Cs, space group: $Cmcm$, no. 63) and similar to the slightly distorted material BaFe$_2$Se$_3$ (space group: $Pnma$, no. 62), as shown in Fig. \ref{fig1}(c)\cite{Hong1972,Reiff1975,Klepp1996,Saparov2011,Du2012,Caron2012,Nambu2012,Wang2016}. The thermal activation gap in BaFe$_2$S$_3$ ($\sim70 meV$) \cite{Reiff1975} is the smallest among the Fe-based ladder compounds, and photoemission studies suggest that both localized and itinerant electrons coexist at room temeprature\cite{Ootsuki2015}. The Fe$X$ ($X=$ Se, S, As, and P) tetrahedra are common among the 1D AF ladder and 2D stripe ordered materials\cite{Lynn2009,Huang2008,Zhao2012,Wangm2014}. However, in contrast to the Fe$X$ tetrahedra in the other 1D AF ladders\cite{Wang2016}, the moments of BaFe$_2$S$_3$ are smaller ($\sim1.2\mu_B$/Fe) and aligned along the rung direction, as shown in Fig. \ref{fig1} (d)\cite{Takahashi2015}, and the distance of the Fe-Fe bonds along the AF direction (leg) is shorter than that along the ferromagnetic (FM) direction (rung). Hence, the spin dynamics, predominately governed by the geometry of the lattice, could be different in BaFe$_2$S$_3$. Accordingly, it is important to measure the spin waves of BaFe$_2$S$_3$ and extract the exchange interactions in order to compare with the other 1D and 2D analogs.

In this paper, we report inelastic neutron scattering (INS) studies on the spin waves of a BaFe$_2$S$_3$ powder sample. Similar to our measurements on RbFe$_2$Se$_3$\cite{Wang2016}, we observe an acoustic branch and an optical branch of spin waves, consistent with two inequivalent irons in the magnetic Brillouin zone. From the spherically averaged spectra on the powder sample, we are able to extract a spin gap, two band tops of the acoustic branch along two directions, and the minimum and maximum energies of the optical branch. By solving the Heisenberg Hamiltonian of the ladder structure with the observed constraints, we determine a set of parameters ($SJ_R=-71\pm4, SJ_L=49\pm3, SJ_2=-15\pm2, SJ_7=3.0\pm0.5$, and $SJ_s=0.1\pm0.04$ meV) with a strong intraladder FM exchange interaction along the rung direction that fits the experimental data well. The results demonstrate that the spin fluctuations are comparable among various parent compounds of the FeSC, while the exchange interactions that are previously proven universal are not unique for the stripe AF ordered parent state of the FeSC.

The BaFe$_2$S$_3$ samples were grown using the Bridgman method\cite{Wangm2014}; they formed in small needle-like single crystals, making them extremely difficult to align. Hence we ground 8 g of the single crystals into a powder for this experiment. Our INS experiment was carried out on the ARCS time-of-flight chopper spectrometer\cite{Abernathy2012} at the Spallation Neutron Source, Oak Ridge National laboratory (SNS, ORNL).  The powder sample was sealed in an aluminum can and loaded into a He top-loading refrigerator. The sample was measured with incident beam energies of $E_i=$ 50, 150, and 250 meV at 5 K. The energy resolutions for these incident beams were $\Delta E=$2.2, 7.0, and 13.3 meV, as determined by the full width at half maximum (FWHM) of the energy cuts at $E=0$ meV.

Figure \ref{fig2} shows INS spectra and cuts for the BaFe$_2$S$_3$ powder samples with different incident energies. In Fig. \ref{fig2}(a), we can see intense excitations at $Q=1.27$ \AA$^{-1}$, dispersive excitations stemming from $Q=2.19$ and 2.81 \AA$^{-1}$, weak excitations at $Q=3.59$ \AA$^{-1}$ and a gap around 5 meV for all the $Q$s. The spectrum resembles the spin waves observed on the ladder compound RbFe$_2$Se$_3$\cite{Wang2016}. The four $Q$s are consistent with the AF wave vectors at $(H, K, L)=(0.5, 0.5, 1), (2.5, 0.5, 1), (3.5, 0.5, 1)$, and (0.5, 0.5, 3), revealing that the excitations are the spin waves of BaFe$_2$S$_3$. Here, $(H, K, L)$ are Miller indices for the momentum transfer $|Q|=2\pi\sqrt{(H/a)^2+(K/b)^2+(L/c)^2}$, where the lattice constants are $a=8.79, b=11.23$, and $c=5.29$ \AA \cite{Takahashi2015}. The flat excitations with intensities increasing with $Q$ below 30 meV are phonons associated with the sample and the thin aluminum can.

To determine the spin gap and dispersion relations quantitatively, we present constant $Q$ cut integrated within $Q=1.27\pm0.1$ \AA$^{-1}$ in Fig. \ref{fig2}(e) and constant energy cuts within $E=6\pm1, 12\pm1, 18\pm1, 24\pm1$, and $30\pm1$ meV in Fig. \ref{fig2}(f). The minimum of the in-ladder plane and out-of-ladder plane spin gaps is $5\pm1$ meV[Supplementary materials]. The spin excitations stemming from $Q=2.19$ and 2.81 \AA$^{-1}$ disperse separately into four peaks with increasing energy. At around 30 meV, the two inner peaks merge together, indicating that the spin waves have reached a maximum along the $[H, 0.5, 1]$ direction. Figures \ref{fig2}(b) and \ref{fig2}(g) present the dispersive spin excitations at $Q=3.59$ \AA$^{-1}$ with $E_i=80$ meV. The spin excitations continuously evolve into dispersionless excitations at 70 meV, as shown in Figs. \ref{fig2}(c) and \ref{fig2}(h). This energy ($\sim70$ meV) is higher than the cut-off energy of phonons and the intensities decrease with inceasing $Q$, indicating that they are magnetic excitations of BaFe$_2$S$_3$. The dispersion relation at $Q=(0.5, 0.5, 3)=3.59$ \AA$^{-1}$ corresponds to the dispersion along the [0.5, 0.5, L] direction. Thus, the dispersionless spin excitations at 70 meV can be ascribed to the zone boundary excitations along the $L$ direction. Gaussian peak fittings to the constant $Q$ cuts of the dispersionless spin excitations in Fig. \ref{fig2}(h) show centers at $71\sim72$ meV. The energy is significant lower than the observed spin wave maximum ($\sim 190$ meV) along the same direction for RbFe$_2$Se$_3$\cite{Wang2016}.

\begin{figure}[t]
\includegraphics[scale=0.4]{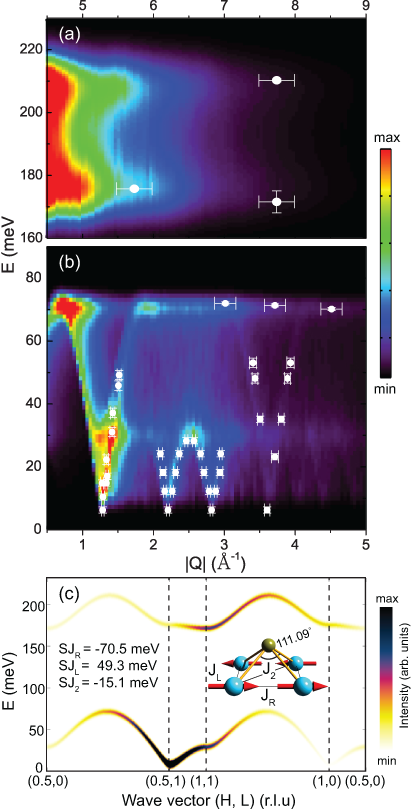}
\caption{ Comparisons between the SpinW simulated spin excitation spectra and experimental determined dispersion relations (white points) for BaFe$_2$S$_3$. (a) Instrumental resolutions of 13.3 meV and (b) 5 meV have been convolved for comparison with the experimental data in Figs. \ref{fig2}(a)-\ref{fig2}(d). (c) SpinW simulated spin excitations along high symmetry directions in the [H, L] 2D Brillouin zone for the parameters labeled on the figures. The other parameters, $SJ_7$ and $SJ_s$, have been fixed at 3.0 and 0.1 meV, respectively. The color represents intensities. We convolve a constant 5 meV instrumental resolution for visualization. The inset in panel (c) shows a tetrahedron and associated exchange interactions.}
\label{fig3}
\end{figure}

In Fig. \ref{fig2}(d), we present the optical spin waves measured with $E_i=250$ meV at 5 K. Two flat branches of excitations are observed. The center of the lower branch is determined to be at $171.6\pm0.3$ meV within $Q=5.75\pm0.25$ \AA$^{-1}$ and $176\pm2$ meV within $Q=7.75\pm0.25$ \AA$^{-1}$, and that of the higher branch is at $210.7\pm0.3$ meV within $Q=7.75\pm0.25$ \AA$^{-1}$. The low and high branches of magnetic excitations are consistent with them being the minimum and maximum of the optical branch of the spin waves of BaFe$_2$S$_3$. The extracted spin wave dispersion relations have been plotted in Fig. \ref{fig3}.

BaFe$_2$S$_3$ at ambient pressure exhibits a stripe ordered structure similar to that of RbFe$_2$Se$_3$\cite{Wang2016}. We proceed to employ the Heisenberg Hamiltonian that has been used to describe the spin waves of the ladder compound RbFe$_2$Se$_3$ and other 2D stripe systems to fit the dispersion relations and extract the magnetic exchange interactions for BaFe$_2$S$_3$\cite{Zhao2009,Ewings2011,Harriger2011,Wang2015, Wang2016,Zhao2014}. The spin Hamiltonian is written as
 \begin{equation} 
  \hat{H}=\sum_{r,r^\prime}\frac{J_{r,r^\prime}}{2}\bf{S}_r\cdot S_{r^\prime}-\it{J_s}\sum_r(\bf{S}_{r}^z)^2,
  \label{eq1}
 \end{equation}
where $J_{r,r^\prime}$ are the effective exchange couplings and $(r, r^\prime)$ label the iron sites, $J_s$ is the single ion Ising anisotropy term\cite{Yao2010}. By solving Eq. \ref{eq1} using the linear spin wave approximation, the dispersion relations and extrema values can be obtained\cite{Yao2010}. Because we have assumed identical Hamiltonians for the spin waves of BaFe$_2$S$_3$ and RbFe$_2$Se$_3$, the solutions have the same analytical expressions\cite{Wang2016}. The spin gap $\Delta_s$, the tops of the acoustic mode along the $H$ direction ($E_{1t}^{H}$) and $L$ direction ($E_{1t}^{L}$), and the bottom ($E_{2b}$) and top ($E_{2t}$) of the optical mode are as follows:
 \begin{equation} 
 \begin{split}
 & \Delta_s=2S\sqrt{J_s(2J_L+2J_2+J_7+J_s)} , \\
 & E_{1t}^{H}=2S\sqrt{(2J_L+2J_2+J_s)(J_7+J_s)} ,\\
 & E_{1t}^{L}=2S\sqrt{(J_L+J_2+J_s)(J_L+J_2+J_7+J_s)} ,\\
 & E_{2b}=2S\sqrt{(2J_L-J_R+J_s)(2J_2-J_R+J_7+J_s)} ,\\
 & E_{2t}=2S\sqrt{(J_L-J_R+J_2+J_s)(J_L-J_R+J_2+J_7+J_s)} .\\
  \label{eq2}
  \end{split}
 \end{equation}
 The $J_R, J_L$ and $J_2$ are the intraladder exchange interactions along the rung, leg, and diagonal directions, respectively. $J_7$ is the seventh nearest neighbor (NN) exchange interaction of irons between two ladders, as defined in Fig. \ref{fig1}(c). The expressions in Eq. (\ref{eq2}) correspond to the wave vectors at $Q=(H, L)=(0.5, 1), (1, 1), (0.5, 0.5), (1, 1)$, and $(1, 0.5)$, respectively. The $K$ for these wave vectors is 0.5.

From the spherically averaged INS data, we have determined the values for these extrema, where $\Delta_s\approx5, E_{1t}^{H}\approx30, E_{1t}^{L}\approx72, E_{2b}\approx172$, and $E_{2t}\approx211$ meV. Solving Eq. (\ref{eq2}) would lead to two sets of mathematical solutions. By comparing with the experimental data, the two sets of parameters are determined as $SJ_L=49.3, SJ_2=-15.1$ meV and $SJ_L=-14.3, SJ_2=48.4$ meV, respectively, while the other interactions, $SJ_R=-70.5, SJ_7=3.0$, and $SJ_s=0.1$ meV, are the same. The two sets of parameters fit our spherically averaged data equally well. However, there is a difference for the optical spin wave branch for single crystals[Supplementary materials]. The intensity distribution of the optical mode for the second set of parameters disagrees with that of RbFe$_2$Se$_3$,  where the intensities at $(H, L)=(1, 1)$ are stronger than that at $(1, 0)$\cite{Wang2016}. The FM $J_L$ is also contrary to a first-principles' calculation\cite{Suzuki2015}, which predicts an AF $J_L$ and a FM $J_R$. Furthermore, the inferred $SJ_2=48.4$ meV is much larger than the expectation for a superexchange interaction between two irons with the distance of 3.78 \AA\cite{Zhao2014,Wang2016}. Thus, the second set of parameters is unlikely to be a physical solution for the Hamiltonian for the spin waves of BaFe$_2$S$_3$. 

We hence determine the products of the spin $S$ and exchange interactions as $SJ_R=-71\pm4, SJ_L=49\pm3, SJ_2=-15\pm2, SJ_7=3.0\pm0.5$, and $SJ_s=0.1\pm0.04$ meV for BaFe$_2$S$_3$. The errors are estimated by considering the effects to the spin wave extrema in Eq. (\ref{eq2}). There should be other weak out-of-ladder plane exchange couplings, e.g., $J_5$, that give rise to the three dimensional magnetic order, as shown in Fig. \ref{fig1}(a). However, we could not determine them from the spherically averaged powder data. We used the SPINW program\cite{Toth2015}, which employs classical Monte Carlo simulations and linear spin wave theory, for simulations. The simulated spherically averaged spectra based on the determined exchange interactions together with the dispersion relations extracted from our experimental data and the spin wave spectrum for single crystals are plotted in Fig. \ref{fig3}. The simulated spectra match the experimental data very well.

\begin{table}[t]
\caption{The magnetic exchange couplings and NN Fe-Fe distances along the antiferromagnetic ($J_{AF}$ and $d_{AF}$) and ferromagnetic ($J_F$ and $d_F$) direction, respectively, and the exchange couplings along the diagonal direction for various Fe-based materials\cite{Zhao2009,Ewings2011,Harriger2011,Wang2015,Wang2016}. The bond distances, $d_{AF}$ and $d_F$, are in unites of angstrom (\AA).}
\begin{tabular}{ccccccc}
\hline \hline
Compounds            & $SJ_{AF}$   & $SJ_{F}$  & $SJ_2$ (meV) & $d_{AF}$   & & $d_{F}$   \\ \hline
CaFe$_2$As$_2$   & $50\pm10$      & $-6\pm5$      & $19\pm4$   & 2.753 & $>$& 2.703 \\
BaFe$_2$As$_2$   & $59\pm2$      & $-9\pm2$      & $14\pm1$   & 2.808  &$>$ &2.786 \\
SrFe$_2$As$_2$   & $39\pm2$      & $-5\pm5$      & $27\pm1$   & 2.785  & $>$&2.756 \\
Rb$_2$Fe$_3$S$_4$ & $42\pm5$      & $-20\pm2$    & $17\pm2$     & 2.76  & $>$& 2.70    \\ 
RbFe$_2$Se$_3$ & $70\pm5$      & $-12\pm2$    & $25\pm5$     & 2.77  &$>$ & 2.64    \\ 
BaFe$_2$S$_3$ & $49\pm3$      & $-71\pm4$    & $-15\pm1$     & 2.64  &$<$ & 2.70    \\ \hline \hline
\end{tabular}
\label{table:t1}
\end{table}

 We list the known $SJ$'s and Fe-Fe bond lengths for the stripe ordered iron-based materials in Table \ref{table:t1}. The magnitude of the AF $SJ_L$ for BaFe$_2$S$_3$ is comparable with the $SJ_{AF}$ along the AF ordered direction for the other 1D and 2D analogs\cite{Zhao2009,Ewings2011,Harriger2011,Wang2015, Wang2016}. However, the strong FM $J_R$ and FM $J_2$ are distinguishable. Following the Goodenough-Kanamori rules\cite{Goodenough2008,Pavarini2012}, a superexchange interaction ($J_2$) connects $d$-orbitals of two magnetic atoms ($M$) via $p$-orbitals of the atom ($X$) in-between. For the case of an $M-X-M$ angle $\alpha=180^\circ$,  both the $d$-orbitals couple to the same $p$-orbital, resulting in an AF $J_2$. However, for the angle $\alpha=90^\circ$, the $d$-orbitals couple to two orthogonal $p$-orbitals, making it impossible for an electron on one $d$-orbital to reach the $d$-orbital on the other site. In this case, the superexchange mediated via the Coulomb exchange on the connected two orthogonal $p$-orbitals is expected to be ferromagnetic. In BaFe$_2$S$_3$, the angle of the Fe-S-Fe along the diagonal direction is 111.09$^\circ$. The competition between the AF and FM superexchange processes could give rise to a FM J$_2$. The extracted FM $J_2$ and the strong FM $J_R$ in BaFe$_2$S$_3$ could be ascribed to this unique geometry, where the bonds along the AF ordered direction ($d_{AF}$) are shorter than that along the FM ordered direction ($d_F$), as shown in Table \ref{table:t1}. The diagonal direction always leans towards the stronger NN exchange ($J_1$) direction and $J_2$ also exhibits the same sign as $J_1$. The $J_2$ could be dominated by the closer $J_1$ in the stripe AF ordered Fe-based materials. On the other hand, the presence of a possible biquadratic exchange interaction could also account for the effective FM $J_2$\cite{Luo2016}. Interestingly, a direct fitting with the $J_1-J_2$ model to the spin waves of La$_2$CuO$_4$ also results in a FM $J_2$, which has been ascribed to the effect of a cyclic or ring exchange interaction\cite{Coldea2001}.
 
The ratio of the exchange interactions has been suggested to be crucial for the SC pairing symmetry and even whether the superconductivity occurs in the FeSC\cite{Yang2013}. A possible orbital ordering transition near 200 K merges gradually together with the magnetic ordering transition ($\sim120$ K) at 2 GPa, accompanying with the abrupt increases of $T_N$ and the moment sizes at 1 GPa\cite{Yamauchi2015,Chi2016}. The superconductivity emerges around 10 GPa, where the magnetic order has been suppressed\cite{Takahashi2015,Yamauchi2015}. Clearly, the orbital ordering, magnetism, and superconductivity are strongly coupled and all of them are sensitive to pressure. The exchange interactions we extract from BaFe$_2$S$_3$ should be related to its unique FeS tetrahedra. Our results clearly are important for any theoretical modeling of the superconductivity based on the spin fluctuation mediated mechanism and for any theoretical investigation of the interplay between the magnetic ordering, orbital ordering, and superconductivity. 

In summary, we have measured the spin wave spectra of the stripe AF order in the 1D ladder compound BaFe$_2$S$_3$ on a powder sample. Guided by the analytical expressions for the extrema of the spin waves and their experimentally determined values, the exchange interactions have been successfully determined. Spherically averaged simulations using the parameters so-determined match well the measured spectra. The explicit values for the exchange interactions in BaFe$_2$S$_3$ are distinct from those of the other 1D and 2D analogs due to its unique structural geometry. The results reveal that the 1D AF ordered ladder parent state of the superconductivity in BaFe$_2$S$_3$ exhibits the commonly observed antiferromagnetic spin excitations just as in the parent compounds of the other FeSC. However, there are important quantitative differences from the previously realized combinations of exchange interactions for the stripe AF ordered parent state of the FeSC, suggesting that a wider range of interactions may still result in superconductivity.

This work was supported by the Office of Science, Office of Basic Energy Sciences, Materials Sciences and Engineering Division, of the U.S. Department of Energy under Contract No. DE-AC02-05-CH11231 within the Quantum Materials Program (KC2202) and the Office of Basic Energy Sciences U.S. DOE Grant No. DE-AC03-76SF008. The research at Sun Yat-Sen University was supported by NBRPC-2012CB821400, NSFC-11275279, NSFC-11574404, and NSFG-2015A030313176. HCJ was supported by the Department of Energy, Office of Science, Basic Energy Sciences, Materials Sciences and Engineering Division, under Contract DE-AC02-76SF00515. H. Luo was supported by NSFC-11374011, NSFC-11674372  and the Youth Innovation Promotion Association of CAS (No. 2016004). The experiment at Oak Ridge National Laboratory's Spallation Neutron Source was sponsored by the Scientific User Facilities Division, Office of Basic Energy Sciences, U.S. Department of Energy.



\bibliography{mengbib}

\end{document}